# Next Steps for Human-Centered Generative AI

Xiang 'Anthony' Chen
UCLA HCI Research
xac@ucla.edu

Jeff Burke
UCLA REMAP
jburke@remap.ucla.edu

Ruofei Du
Google Research
ruofei@google.com

Matthew K. Hong
Toyota Research Institute
matt.hong@tri.global

Jennifer Jacobs
UCSB
jmjacobs@ucsb.edu

Philippe Laban
Salesforce Research
plaban@salesforce.com

Dingzeyu Li
Adobe Research
dinli@adobe.com

Nanyun Peng
Computer Science Department, UCLA
violetpeng@cs.ucla.edu

Karl D.D. Willis
Autodesk Research
karl.willis@autodesk.com

Chien-Sheng Wu
Salesforce Research
wu.jason@salesforce.com

Bolei Zhou
Computer Science Department, UCLA
bolei@cs.ucla.edu

## ABSTRACT

Through iterative, cross-disciplinary discussions, we define and propose next-steps for Human-centered Generative AI (HGAI). We contribute a comprehensive research agenda that lays out future directions of Generative AI spanning three levels: aligning with human values; assimilating human intents; and augmenting human abilities. By identifying these next-steps, we intend to draw interdisciplinary research teams to pursue a coherent set of emergent ideas in HGAI, focusing on their interested topics while maintaining a coherent big picture of the future work landscape.

## 1 INTRODUCTION

The recent development of Generative AI—ranging from large language models [26, 148] to visual generation techniques [100, 115, 157]—promises to revolutionize how humans work in a wide range of tasks [109]. Meanwhile, various research communities, will soon, if not already, be working on topics related to Generative AI.

Historically, when new technological breakthroughs emerged, there was a tendency for researchers in adjacent fields to pursue "low-hanging fruits" and rapidly produce results. While such an approach does accelerate our knowledge of the new technology in the short-term, it somewhat limits researchers' vision from seeing a holistic picture and how certain problems can and should be tackled with cross-disciplinary efforts.

To establish a unified framework that ties various emergent research across disciplines, this paper proposes Human-centered Generative AI (HGAI, pronounced 'H'-/ɡaɪ/) as an overarching topic and lays out specific next steps for achieving HGAI. Our focus is on identifying joint HGAI research opportunities across related technical disciplines, mainly including technical human-computer interaction (HCI) research [73], machine learning, natural language processing, computer vision, and computer graphics[1].

To formulate the definition of HGAI and the next steps, we adopted the process used in analogous prior work [105] and followed a bottom-up approach to conduct three iterations[2] of discussions amongst authors from the above-mentioned disciplines across academia and industry: *(i)* individual brainstorming discussions between the first and every other author; *(ii)* paired deep-dive discussions, each moderated by the first author and involving two other authors from different disciplines; and *(iii)* virtual "walk-the-wall" where every author could see and contribute to all the HGAI next-step ideas represented as an affinity diagram.

Our main contribution is a agenda for future technical research on HGAI that unifies ongoing topics as well as less-explored ideas. Embracing perspectives across different technical disciplines, this agenda intends to draw interdisciplinary teams to a comprehensive list of research opportunities on HGAI, identifying their interested topics while maintaining a coherent big picture of the future work landscape. In the remainders of this paper, we detail the process of developing the HGAI agenda, describe our definition of HGAI, and discuss specific proposed next-steps for technical research on HGAI.

---

[1]Admittedly, HGAI is a class of "wicked problem" [116] that intertwines with more than the above disciplines; we chose to focus on a comprehensive (but not exhaustive) subset of HGAI-related areas so that we can best leverage the authors' expertise to conduct in-depth discussions within the scope of a single paper, while leaving uncharted space for future research.

[2]We describe the discussion process in details in the appendix.

## 2 DEFINING HUMAN-CENTERED GENERATIVE AI (HGAI)

To define HGAI, we start with a walkthrough of the key terminologies below.

**What do we mean by "Generative AI"?** Commonly, Generative AI can be broadly defined as computational processes that can generate new data instances, which contrasts with Discriminative AI that aims at distinguishing between different kinds of data instances [3]. Although the definition of Generative AI is fairly general and can date back to early work (*e.g.*, Topology Optimization [118] proposed in the early 90's), the majority of our discussions were concerned with the recent developments in Large Language Models (LLMs) and Large Multimodal Models (LMMs). In the remainders of this paper, our mentioning of Generative AI mainly refers to LLMs and LMMs.

**What do we mean by "Human" in HGAI?** There are various stakeholders involved in the ecosystem of Generative AI, including

(1) people whose data is used for model training (*e.g.*, artists' and designers' work),
(2) people who label and moderate the data (*e.g.*, to filter out toxic contents [5]),
(3) people who develop Generative AI models (*e.g.*, academic professors/students and employees in tech companies who build LLMs),
(4) people who develop systems that use Generative AI models (*e.g.*, game development platforms that use LLMs to enable conversational characters),
(5) end-users of Generative AI and its applications,
(6) and finally, people who are impacted (in)directly by various (un)intended consequences of Generative AI (*e.g.*, teachers grading AI-generated essays).

Our discussion of HGAI primarily focuses on the betterment of individuals who own the data and the end-users of Generative AI, although some of our next-step ideas and calls-for-actions do speak to the other stakeholders as well. For example, appropriately leveraging AI in students' writing process (an example of augmenting human abilities) can lead to more transparent ways for teachers to grade essays.

**What do we mean by "Human-centered Generative AI"?** We propose the following definition and then differentiate it from related concepts in prior work:

The three HGAI objectives rest on a pyramid structure. Aligning with human values is the foundational level—a pre-requisite because no other objectives matter unless we can ensure Generative AI behaves ethically without violating the values of the specific user population it is concerned with. Next, just like how human-human communication is essential to collaboration, in order for Generative AI to augment human abilities, it first needs to assimilate, *i.e.*, capture, understand, and realize, human intents.

**Related concepts in prior work**. Human-centeredness is not unique or specific to AI or Generative AI. In [33], Chancellor surveyed and identified several prior framings and definitions, *e.g.*, distinguishing from work that mainly focused on the technical aspect [68], that "must take account of varied social units that structure work and information" [82], and that asked what should be, rather than could be, produced [61], and that related to the

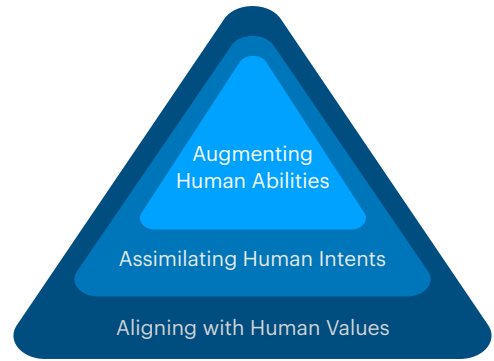

> **Definition**
>
> Human-centered Generative AI (HGAI) should achieve three levels of human-centered objectives: *(i)* Aligning with human values; *(ii)* Assimilating human intents; and *(iii)* Augmenting human abilities.

**Figure 1: Our definition of Human-centered Generative AI (HGAI) across three levels.**

"social-technical gap" [14] in CSCW literature. Building off of such historical contexts, Chancellor then proposed a renewed definition of human-centeredness in machine learning as following a set of practices to achieve balances between technical innovation and human and social concerns. The above evolution of human-centeredness is inline with Level 1 in our definition centered on human values.

Schmidt, in defining "Interactive Human Centered Artificial Intelligence" [122], pointed out the importance of "amplifying the human mind without compromising human values", which is in agreement with our Level 1 and 3.

Another recent paper by Capel and Bereton [30] surveyed how human-centeredness has been interpreted in various subfields, *e.g.*, Explainable and Interpretable AI, Human-Centered Approaches to Design and Evaluate AI, Humans Teaming with AI, and Ethical AI, based on which they proposed a new definition: "Human-Centered Artificial Intelligence utilizes data to empower and enable its human users, while revealing its underlying values, biases, limitations, and the ethics of its data gathering and algorithms to foster ethical, interactive, and contestable use." This definition echos both Level 1 and 3 in our definition. We further include Level 2, which is an important aspect of HGAI as humans need to express and realize their intents to control what contents will be generated. Related to our Level 1 definition of HGAI, a recent talk by Pascale Fung [29] laid out different types of harms caused by LLMs ranked by severity, ranging from offensive and biased language, to deepfake and discriminatory generation, to law-breaking privacy violation and misinformation, and to life-threatening acts such as medical misdiagnosis and terrorism.

On the industry side, various organizations have been conducting alignment research, *e.g.*, OpenAI's approach[10] of "engineering a scalable training signal for very smart AI systems that is aligned with human intent", including training AI systems using human feedback, to assist human evaluation, and to generate explanations of LLMs [21]. Such effort is mostly related to our Level 2 definition of HGAI.

Below we provide an overview of each level of HGAI (Figure 1), each of which necessitates and builds off of previous level(s).

## 2.1 Aligning with Human Values

The foundational objective of HGAI is the alignment with human values. We consider values as the fundamental beliefs that define the ethics of Generative AI. Admittedly, there is no easy way to define a one-size-fits-all value system for the entire humanity[3]; instead, human values often vary across cultures and regions and should be carefully considered with respect to the specific stakeholders of Generative AI.

Alignment with human values has been widely discussed in value-sensitive design [59] and, more recently, in addressing "black box" AI's violation of human values [40, 136]. To align Discriminative AI with human values, one example is preventing racial biases when performing facial recognition [31]; in a similar vein, Generative AI that aligns with human values should not generate racially-biased images of human faces when given certain text prompts. Note that the need for alignment does not mean there exist some universal human values to be aligned with. Generative AI's behaviors need to consider values of specific population groups involved in the model's development and usage and acknowledge that a solution without trade-offs might not exist. For example, Generative AI might align well with small business owners by helping them inexpensively create artworks or slogans for advertisement; yet, in the mean time, such generated contents might have displaced or violated artists' rights to profit from their work, thus misaligning with artists' values.

To ensure HGAI's alignment with human values, simply training a larger model is not enough [110]; we argue that we should follow a human-centered process, not only for designing systems that utilize Generative AI, but further for creating Generative AI models in the first place. Human-centered design is a well-established body of methods to ensure that a system will actually benefit the stakeholders it intends to serve. However, besides benefiting its intended users (*e.g.*, the aforementioned small business owners), Generative AI should also prevent causing harm to affected people (*e.g.*, artists whose work has been incorporated into the model). Thus, a human-centered process for Generative AI should be extended from design activities to the model training and development stages, particularly by involving people who might be negatively impacted by the resultant model. In §3, we discuss various HGAI next-steps throughout this process.

## 2.2 Assimilating Human Intents

Unlike Discriminative AI where the input is some existing information (*e.g.*, an image for object recognition or some text for summarization), input to Generative AI is much more open-ended, making the assimilation of human intents more challenging.

Foremost, HGAI should offer an expressive medium through which users can freely and effectively convey their intents of generating certain contents. Existing approaches, such as text prompts, might be a convenient shortcut to convey intent; however, it remains limited in many scenarios where text alone either cannot clearly represent a user's intent or does not allow a user to iteratively refine their intent expression.

Further, HGAI should ensure that the generative process follows a user's expressed intent. Since the generative model is often trained on massive amounts of data on the Internet, it could produce uncontrollable behaviors that deviate from a user's intent (*e.g.*, hallucinated contents [76]). As such, HGAI should provide explicit control mechanisms for a user to steer the generative process or allow a user to "talk back" to the Generative AI by editing an imperfect result.

Meanwhile, enabling user control of Generative AI is a balance act. The goal is to provide an appropriate amount of control to the end-users: too much control overwhelms users, making Generative AI less accessible, whereas too little control makes the model output unsteerable by the user. Finding the right balance of control in each application setting is important, as demonstrated in some recent work on controlling Generative Adversarial Networks (GANs) [47].

Note that it is not always the case that humans have clear and strong intents that can be articulated in any form (text prompts or otherwise). Sometimes implicit control mechanisms (*e.g.*, learning preferences from past interactions) are important too. HGAI should explore the diversity of control mechanisms that range from implicit to explicit, text to rich media formats, which adapt to human intents in-the-moment because intents might be uncertain, constantly-evolving, and perhaps best supported in a mixed-initiative manner [71].

Finally, it is important to realize that not all human intents are benign (*e.g.*, using LLMs to fabricate false news); therefore, HGAI should foremost align with human values (Level 1) before committing to assimilating human intents (Level 2).

## 2.3 Augmenting Human Abilities

Generative AI that can assimilate human intents should then aim to augment human abilities in achieving their domain-specific goals. Despite the promises of Generative AI, there often remains a gap between what the AI model can generate and how such AI can actually benefit a domain user's work. For example, consider OpenAI's Codex—an LLM capable of generating functional code snippets: such a model alone might not be beneficial to a programmer who works in conventional integrated development environments (IDEs). In contrast, GitHub's Copilot—an LLM with similar code generation capabilities—is fully integrated with programmers' IDEs. One simple example of such integration is allowing programmers to control how AI completes their in-progress programs—generating one line *vs.* multiple lines of code—a feature that is very basic yet considerate of programmers' work practices and goes beyond generating code alone.

To further close the gap, the design of HGAI systems should aim to divide the labor in ways that match what the human and AI each does best. For example, consider video production. Perhaps the human is best as the director who asks GPT to write the script and keep the scene settings and back stories consistent, which is often challenging for human screenwriters. As another example, consider novice users interacting with Generative AI for visual design tasks. As these users are probably unfamiliar with the best choice of terminology in a text prompt, HGAI can start with guiding the user to formulate the scope of what visual contents they want to create, then incrementally brainstorm examples to populate a set of candidates, and then help the user continuously narrow down and refine their choices. Such a three-step workflow can lead to effective use of Generative AI compared to improvising and trying out different prompts.

---

[3]Although there have been multiple ongoing efforts of unification, *e.g.*, the Blueprint for an AI Bill of Rights [4] and PCAST Working Group on Generative AI [11].

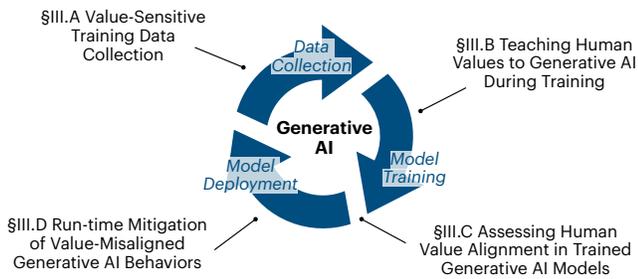

§III.A Value-Sensitive Training Data Collection
§III.B Teaching Human Values to Generative AI During Training
§III.D Run-time Mitigation of Value-Misaligned Generative AI Behaviors
§III.C Assessing Human Value Alignment in Trained Generative AI Models

§III.E Case Study: Mitigating Harm to Human Creators Caused by Generative AI

**Figure 2: Overview of HGAI Level 1: next-steps in aligning with human values.**

## 3 HGAI NEXT-STEPS: ALIGNING WITH HUMAN VALUES

As shown in Figure 2, this section lays out next-steps for HGAI throughout the Generative AI lifecycle, from collecting training data, to training models, to assessing trained models, and to runtime mechanisms. Further, we zero in on a case study of mitigating Generative AI's harm to human creators.

### 3.1 Value-Sensitive Training Data Collection

Since modern AI is mostly data-driven, a main culprit of ethical issues is the training data. For Discriminative AI, some training data might cause AI to aim at the wrong target [108], such as using "income" to determine a credit score because there is no other better attributes, *e.g.*, "credit worthiness". To mitigate such limitations engendered in training data, one approach is simply making the process of collection and the composition of training data as transparent as possible, *e.g.*, via a specification document like a datasheet [62]. For HGAI, the next-steps should foster a value-sensitive training data collection process.

*3.1.1 Preventing training data from giving rise to biases.* Since Generative AI aims to learn the distribution of a certain domain to generate new data instances, it is even more critical to ensure that the training dataset is unbiased. The ever-increasing size of required dataset and its breadth goes well beyond a traditional (non-generative) learning problem, making it even harder to achieve this. Meanwhile, we should recognize that sometimes there is no universally unbiased dataset or solution; therefore, when determining biases, it is important to contextualize a dataset by which population group its resultant model aims to serve. Thus the next-steps of HGAI should aim at the following objectives.

Expanding the well-established user-centered [81] and task-centered [87] design process to encompass the early stage of data collection to ensure no misrepresentation of the affected user populations will propagate to downstream models and systems. Fei-Fei Li mentioned that, before the start of a recent project in her lab to benchmark robotic tasks, their team conducted a large-scale user study to identify the winning task most beneficial to the target users, which became the focus of the project [6]. Based on such user-centered practices, the next-step is to formulate and evaluate a generalizable protocol to guide the training data collection process prior to developing Generative AI.

Rather than suppressing potential biases so the model would never learn such behaviors, another direction for next steps is to develop Generative AI that is aware of biases already existing in the real world. Consider how a text-to-image AI might learn biases and generate predominantly male images for *CEO*. Since such biases stem from the gender imbalance in the executive world, it might be worth for Generative AI to learn about such phenomena, so that the model not only will prevent generating biased images but can further explain to the user how it unbiases the generation or even let the user control such processes.

As a biased Generative AI model's output might permeate into the real world (*e.g.*, people using ChatGPT to write various sorts of documents), it will soon become imperative to stop such biased AI-generated data from being used to train future models. According to Veselovsky *et al.*, 33-46% of crowd workers are estimated to use LLMs in completing their tasks [139]. One next-step is to incorporate ongoing efforts that detect AI-generated contents (*e.g.*, [74, 103]) into the screening of training data, which still remains as one highly challenging topic.

*3.1.2 Preventing creators' data from being used for training Generative AI.* In a recent discussion, Pamela Samuelson described latest legal cases and the challenge of determining what constitutes infringement in the context of AI-generated contents [42]. Similar to how the General Data Protection Regulation (GDPR) influenced a wide range of changes in how technology handles user data (*e.g.*, cookies on websites), we can anticipate changes in Generative AI systems as new advances on the legal front take place, such as adding Adversarial noises [123] or "certified" watermarks [13, 17, 57] to prevent unobstructed usage of artists' work. Next-steps for HGAI are as follows.

One grand challenge is enabling creators to protect their works in public domains from being used for training Generative AI, which requires the entire industry and research community to establish new protocols of data collection. For example, as of Dec 2023, the Adobe Firefly generative AI model is claimed to be trained on a dataset of licensed content, such as Adobe Stock, and public domain content where copyright has expired[4]. Similar to how open source licenses enable and regulate how one's code can be used, we need to develop similar mechanisms for writers, artists, and designers to specify permitted usage of their work. For example, one possible mechanism is that artists who opt-in to contribute their work for training a model can have access to that model's generated contents for their future projects.

---
[4]https://www.adobe.com/products/firefly.html#faq

Alternatively, Generative AI developers should allow creators to audit existing training data with tools to help them identify whether their work has been inadvertently included. For example, "Have I Been Trained"[5] made an important step to help artists detect whether their work is in public datasets like LAION-5B[6].

## 3.2 Teaching Human Values to Generative AI During Training

Beyond learning from labeled data, prior work has discussed and demonstrated interactive machine learning approaches [56] of teaching Discriminative AI models [129, 156]. However, teaching machines beyond labels remains a rather under-investigated problem and below we discuss some starting points of teaching human values to Generative AI.

*3.2.1 Defining value-sensitive metrics and reward functions.* Researchers have found that western, educated, industrialized, rich, and democratic (WEIRD) populations have been dominating the participant groups in behavior science [69] and HCI [92]. A similar issue also exists in some Generative AI models that perform unequally. For example, it has been shown that using ChatGPT as a Question-Answer tool works to various degree [133]: the success rate seems higher for users in developed countries than those in developing countries, *e.g.*, asking for a recipe for making Western *vs.* non-Western cuisines. Such inequality in performance causes a vicious cycle: populations who benefit less from Generative AI will become less engaged and contribute less, resulting in future models to under-serve them even more severely. Although it is possible to mitigate such inequality via fine-tuning large base models with imitation data, a recent paper has found such approaches still fall short in closing the gap of what is unsupported by the base models [66]. Some next-steps for HGAI are as follows.

We should study how Generative AI developers currently are aware of and how they overcome the performance inequality issues. For example, one recently-proposed approach is Constitutional AI [16]—the development of AI that complies with explicitly written human rights and ethical principles, such as privacy, transparency, accountability, and non-discrimination. It would be interesting to learn how such an approach works in practice amongst Generative AI developers.

Rather than assuming there is an invisible, one-size-fits-all objective function to define human values, we should introduce population-specific metrics in the loop of training Generative AI models[7]. Via a collaboration between disciplines, we can approach this goal from both a software engineering perspective (how to efficiently build multiple generative models tailored to specific populations?) and a machine learning methodological perspective (is it possible to incorporate multiple population-specific objectives into the development of one model?)

One opportunity is to work with social scientists to develop an ontology to better define, categorize, and quantify ethics and value related issues in Generative AI. Such an ontology will inform HGAI research with a comprehensive and hierarchical view, based on which we can better conduct systematic studies, targeted data collections, and develop mitigation methods to address ethical issues in Generative AI. Using ontology as a tool, AI researchers can better identify more subtle fairness issues [25, 127] and more accurately measure the extent to which the models are biased.

*3.2.2 Adding controls to Generative AI.* Knowing what biases already exist in Generative AI, we can devise targeted controllable generation methods to mitigate such biases, *e.g.*, by inducing negative biases and positive biases for another demographic, or by equalizing biases between demographics [126]. As another example, using a constrained decoding technique, it is possible to limit the generation of Ad Hominem language that targets some features of a person's character instead of the position the person is maintaining [127].

Currently, such controls are administered as part of the model training process and future work can develop new interfaces to let end-users interactively manipulate such controls. For example, we can enable end-users to interactively reduce biases in images created by Generated Adversarial Networks, *e.g.*, by selecting additional images to balance the proportion between genders or adjusting weights assigned to images [55].

## 3.3 Assessing Human Value Alignment in Trained Generative AI Models

Despite extensive testing in the lab, Generative AI deployed to the real world still likely to behave suboptimally and unexpectedly; as such, next-steps for HGAI can perform risk assessment or auditing of models post-training.

*3.3.1 Risk assessment of Generative AI models.* After a Generative AI model is developed, there remains important work of risk assessment for each application that utilizes the model, trying to foresee its ethics-related impact. Performing such risk assessments could be resource-demanding and there are no standardized approaches at present.

One next-step could be research and studies of risk assessment methods. For example, low-risk applications can employ checklists to observe how much the generated contents violate some pre-defined rules. High-risk applications may require costly virtual sandboxing experiments to contain possible riskier actions before public release, such as generative approaches to perform medical diagnoses or controlling field robots.

Inspired by some recent work on generative agents for simulating human behavior [112], another next-step is to develop toolkits that support custom simulative experiments with generative AI in virtual environments to elicit possible problematic behaviors.

*3.3.2 Auditing Generative AI models.* As Generative AI is frequently updated with significant changes, it is important to perform repeated timely assessment. One existing solution is integrating assessment capabilities with the model development, such as a tool suite built on top of TensorFlow Model Analysis that can be used to compute and visualize commonly-identified fairness metrics for classification models, *e.g.*, false positive and negative rates [1]. Further development of automatic auditing can alleviate the long-term workload of having to periodically reassess an evolving model.

---

[5]Have I Been Trained: https://haveibeentrained.com
[6]LAION-5B: https://laion.ai/blog/laion-5b
[7]Admittedly, population-specific metrics can be a double-edged sword if being used by, say, extreme groups that represent dangerous ideology. To prevent this, there should be mechanisms to detect AI with problematic behavior, *e.g.*, generating hate speech.

One related approach is auditing algorithms—"a method of repeatedly querying an algorithm and observing its output in order to draw conclusions about the algorithm's opaque inner workings and possible external impact" [99]. One recent example of this approach is polling different demographic groups to measure how language models underrepresent or misalign with them [119]. Generative AI presents unique challenges to perform such audits due to the sheer amount and variety of generated contents. Further, it is often unclear how to track or assess whether certain changes in the Generative AI result in better or worse contents, such as writing or artwork where the assessment can be subjective and unscalable if requiring human involvement. To make it possible to audit generative AI, there are several next-steps as follows.

Collecting datasets of changing generated contents due to model updates and performing analyses to identify unexpected changes, thus the need to perform audits. In the textual domains, *e.g.*, news, some past benchmarks might become part of the future training set, thus it is important to prevent such overlaps along the time axis. In the visual domains, the main challenge to be addressed is defining metrics to effectively and efficiently measure significant changes.

Conducting studies to understand how Generative AI developers and end-users currently are aware of and cope with model updates and changes. Such work requires longitudinal studies or building online community support for users to report unexpected changes over time.

Developing toolkits—both developers and end-users facing—to support auditing (*e.g.*, the AuditNLG library [2]), including curating a set of benchmarks, defining criteria, reporting audit outcomes, and troubleshooting unexpected changes (*e.g.*, certain types of generated contents start to show biases). Such tools can build on recent work that shows the possibilities of using language model automate evaluations over time to track changing behaviors [113].

## 3.4 Run-time Mitigation of Value-Misaligned Generative AI Behaviors

Even after a model is deployed to an application and in the end-users' hands, there are still opportunities to put more guard rails at run-time on Generative AI's behaviors. Admittedly, these solutions simply mitigate inappropriate model behaviors but do not fundamentally correct such models that violate human values (*e.g.*, trained with biased data).

*3.4.1 Informing users of possible unethical behavior.* Model card—a document that details the model's intended use, the data it was trained on, its accuracy, and potential biases—has been a popular approach to inform the public of an AI model's performance and potential limitations, and to increase transparency and accountability in AI development. To support the creation of model cards, there have been toolkits integrated with the AI development pipeline, such as Google's Model Card Toolkit [7]. To further broaden engagement in the community, the Model Card Authoring Toolkit provides a tool that helps members of a community to review and choose from a range of machine learning models based on their shared values, by providing assistance in understanding, navigating, and evaluating the models [124].

For Generative AI, at present, there is a lack of discussion about how model cards should be different than Discriminative AI's. Thus one next-step is to perform a bottom-up study of developers' existing approaches of creating Generative AI model cards, engage other stakeholders (*e.g.*, content creators and end-users) to elicit their feedback on such model cards, and formulate metrics and standards to guide the best practices. One unique challenge to tackle is that documenting Generative AI's problematic output: how to provide a set of samples with sufficient coverage without overwhelming the readers?

At the UI level, future work can explore connecting specific generated examples to sections of the model card. For example, the DALL•E Mini model card describes one of the possible biases as "When the model generates images with people in them, it tends to output people who we perceive to be white, while people of color are underrepresented." [49] This model card section can automatically emerge on the UI upon detecting the intent of generating faces.

*3.4.2 Explainable Generative AI.* Spearheaded by DARPA's initiative [67], a plethora of research has thrived on eXplainable AI (XAI), most of which assumes the context of Discriminative AI that outputs decisions rather than contents. As such, there has been scarce discussion of how to define and enable explainability for Generative AI. One recent work focused on code generation and took a scenario-based approach to elicit developers' needs for explanation when using Generative AI in various programming scenarios: natural language to code, code translation, and code auto-completion [134]. Future work on explainable Generative AI can start with identifying specific aspects of existing XAI techniques that can be transferred or adopted in the generative scenarios. Some next-steps for HGAI are as follows.

Conducting formative studies to understand end-users' needs for explaining Generative AI: when they need explanations and how they act on explanations? For examples, one hypothesis might be the need for explanation when certain prompts do not result in desired generated contents; a successful explanation, in turn, should enable a user to improve their prompts, *i.e.*, higher satisfaction with the new generated contents.

Studying whether and how existing XAI techniques apply to Generative AI and identifying the gap. Based on numerous taxonomies of XAI [131], we can attempt to draw analogies to the generative domains, *e.g.*, how counterfactual techniques [45] can be redefined in the scenario of explaining text-to-image prompts.

Complementary to the above approach that starts from XAI literature, we can also explore techniques that explain the output of Generative AI via participatory design and technology probe [75], spanning multiple generated modalities *e.g.*, text, image, and audio.

Implementing and evaluating explanation techniques in the contexts of representative application scenarios, *e.g.*, programming, writing, and visual design. Similar to how plug-ins enable ChatGPT to access specific domain information, future Generative AI interfaces can provide explanation plug-ins to promote a wide range of available techniques at end-users' disposal.

*3.4.3 Detecting and disabling inappropriate generated contents.* Given how multiple generative AI models might generate the same types of inappropriate contents (*e.g.*, images that contain gender stereotypes), one next-step for HGAI is to enable users to detect such inappropriate contents, specifically:

For models that support such detection, we can employ existing classifiers and leverage LLMs to self-detect consistency to human value (*e.g.*, whether the just-generated text contains toxic contents [142]).

For integrating such detection into the user interface, we should carefully study the impact of detection models' performance: in particular, false negatives might cause users to inadvertently use inappropriate generated contents in downstream tasks.

Another important challenge is to ensure that such detection models align with end-users' values, such as allowing them to program-by-example and specify what contents should be considered inappropriate.

One next-step following detection is allowing users to disable inappropriate or undesirable content generation, such as disabling the chatbot from talking about politics [8] or disabling age-inappropriate elements when generating stories for children.

*3.4.4 Augmenting input and filtering output.* Given how it is unrealistic to change or even just fine-tune a model at run-time, next-steps for HGAI can instead focus on augmenting the input and filtering output to achieve value-aligned generated contents.

Analogous to how data augmentation [137] can artificially increase the size and diversity of a dataset, one next-step is to develop techniques to augment a user's input to Generative AI (*e.g.*, text prompts) to preempt the generation of biased contents. For example, by inferring from a user's prompt that race could be a latent biased variable, we can append additional terms that request more racially-diverse output. Then, as the model returns a large number of results, we can present a subset of randomly-sampled results to the user.

Another type of output filtering might also mitigate the issue of appropriating other creators' work. Similar to how generating pseudocode can inform a user to implement certain program without directly using others' code, one next-step could be generating 'pseudo artwork' or 'pseudo writing' that represents a creator's style without appropriating their work. Future work should carefully validate this approach of generating styles by closely engaging creators, *e.g.*, via participatory design of how styles are represented and used in end-user applications. There are two key considerations here to pursue such techniques: *(i)* keeping creators in the loop—we should learn from creators what constitute a good piece of 'pseudo work'; and *(ii)* keeping users engaged—a system should provide sufficient tool support for a user to create their own versions of an artist's or a writer's work by following and mimicking their 'pseudo work'.

Augmenting input and filtering output can also address new value-sensitive issues that arise unanticipated by a trained model, *e.g.*, due to the changes of expansion of the user population calling for the inclusion of additional values when using Generative AI.

Meanwhile, it is also necessary to address the trade-offs of the above approaches, such as increased system latency. The system should also make it transparent in how it augments the user's input or filters the model's output, and further let the user have control over such mechanisms, such as setting the number of requested samples to manage latency.

## 3.5 Case Study: Mitigating Harm to Human Creators Caused by Generative AI

To end our discussion of HGAI Level 1, we focus on a case study to address one of the most concerning issues of Generative AI: how it causes harm to human creators, *e.g.*, artists and designers. It has been widely recognized that creators' work in public domains were being scrapped for training Generative AI, causing various kinds of harm to both individuals and the community as a whole [9]. Although our case study here mainly focuses on artists and designers to allow for a deep discussion, other creators are also affected in similar ways, *e.g.*, software engineers whose open-sourced code was scraped by Generative AI and used inappropriately elsewhere.

*3.5.1 Attributing elements of generated contents to the work of creators'.* Imitating, appropriating, or somehow building off of each other's work is not a new phenomenon unique to Generative AI. Human creators often borrow, adapt, and appropriate each others' work in ways that are constructive and acceptable in each specific community. One example HGAI can learn from is the reuse and repurpose of code in open-source software communities. However, unlike source code that structurally follows some programming languages, other generative domains, such as painting and music composition, are much less structured, making it much harder to enable content attribution. As such, some next-steps for HGAI include the following topics.

Analogous to how NLP models summarize text, we can attempt to develop both abstractive and extractive attribution models: the former aims to provide a high-level description of how the generated contents overall can be attributed to certain creators' work while the latter highlight specific elements and map them to certain creators' example work to indicate attribution;

From a user interface perspective, we should also study the ambiguity and scalability of the above approach. *(i)* Ambiguity: how to visualize AI-generated contents being partially influenced by a creator's style or a combination of multiple creators'? *(ii)* Scalability: suppose AI-generated contents mimic a large number of creators' work, *e.g.*, a 'remix' on Spotify that merge numerous musicians' work, is it still useful to show attribution and how to avoid overwhelming end-users?

One natural next-step built on the two above is allowing end-users to remove certain elements from AI's generated contents to avoid imitating other creators' work. We can provide explicit controls for end-users to limit what AI can generate or provide tool support to let them develop their own style in lieu of some creators'.

*3.5.2 Involving creators in the process of developing Generative AI systems.* Despite the harm that has already been done, it seems highly likely that Generative AI will continue to play a major role in the art and design community. As such, some next-steps for HGAI should aim to allow for symbiotic co-existence between Generative AI and human creators.

Going beyond how traditional user-centered design methods ensure a system design provides values to end-users, we should employ similar participatory design methods that also involve creators so that a Generative AI system can both provide values to end-users while minimizing harm done to the creators.

Revisiting value-sensitive design [59] as the values and incentives of the HCI and AI community are probably very different from the values of professional artists. Building less harmful artist-oriented AI technologies requires broadening or redefining our value sets within the HGAI community.

If the intended end-user of a Generative AI system is the creator, we should explore interface and interaction designs that go beyond prompting. Most generative models take text or RGB pixels as input, likely due to technical convenience. However, artists might possess a much larger set of creative methods, *e.g.*, brush strokes, vocals, camera framing, and poetry, and there needs to be a deeper understanding in how Generative AI can support such idiosyncratic creation process.

*3.5.3 Providing an educational platform for creators to explore usages of Generative AI.* On the positive side, AI holds promise to help designers and artists to automate the tedious and repetitive parts of their job. For example, well before robust background removal AI, background or "green screen" removal involves extensive human labor from video artists and editors. Later, as the technology matured, artists and editors who embraced AI tools and learned their pros and cons became more efficient at their work.

As recent generative AI has brought forth lots of new capabilities, creators might find it challenging to catch up with the fast pace. Thus one next-step for HGAI is to provide an educational platform for creators to demystify Generative AI (*i.e.*, understanding the basic principles and current limitations) and explore how to best integrate it into their work.

Exploring the optimal way to collaborate with Generative AI should not be left to each individual human creator; rather, we should study the best practices (*e.g.*, from artists who are also familiar with Generative AI) and provide tool support for the less tech-savvy creators, *e.g.*, tracking and comparing how their work evolves in the course of invoking Generative AI's assistance.

*3.5.4 Ensuring Generative AI developers follow guidelines aimed at protecting designers/artists.* Further, even if we provide guidelines for human-centered methods of AI development for art and design, it remains unclear whether and how researchers, independent developers, and the open source AI community would follow these guidelines. Some next-steps for HGAI include the following topics.

Starting with our own academic communities, we can attempt to set up norms and guidelines that prescribe ways in which researchers might consider using (or not using) training data that contains creators' work. We should discuss the degree to which efforts to assess risk and impact of using Generative AI should be documented in research publications or judged in peer review, similar to how NeurIPS 2020 started asking authors to include a section that discusses the broader impact of their work. Perhaps beyond just a statement alone, research that claims to develop Generative AI tools to serve creators should be judged by whether and how there are partnerships with communities of artists as a component of the project.

Once the above guidelines mature in the research communities, we should aim to extend them to the developer communities. Given how it has become so much easier for individual or hobbyist programmers to fine-tune Generative AI and build their own applications, we should conduct studies to understand their current practices and whether and how developers would follow such guidelines.

Building on the aforementioned educational platforms, we can extend the scope to build and study an online community that allows creators and developers to better communicate their work, respectively. Guidelines for development can be embodied in such creator-developer communication. Developers can get inspirations from creators what Generative AI tools are interesting to build, creators can guide developers to collect training data, and the tools can be developed and tested via a closed-loop collaboration between the two groups.

While the proposed efforts to prescribe guidelines and approaches for reducing harm are important, we need to acknowledge the reality that Generative AI has already caused harm to many professional artists and quite likely presents an existential threat to entire artistic professions, such as illustration and graphic design. Thus it is likely that creators in the art and design community might have already formed a reasonable skepticism when some Generative AI tools promise to benefit individual artists. Overcoming such established aversion is an indispensable part of the next-steps for HGAI. Further, learning from some recent reflection on empathic approaches in accessibility research [19], it is important not to assume or oversimplify the need or accomplishment of engaging with creators. Following best practices that promote benefits while reducing harm to creators should be implemented and assessed as rigorously as the research or development of models. For example, for academic authors, an impact statement is perhaps more appropriately included in the limitation section, acknowledging what has actually been done to mitigate harm, whether such measures are evaluated, and limitations of the effect.

## 4 HGAI NEXT-STEPS: ASSIMILATING HUMAN INTENTS

Perhaps the expression of intent is Generative AI's most distinguishing factor from Discriminative AI. The problem is that intent is ambiguous. Some might argue that the popular approach of text-to-generated-contents already works quite well as it successfully mimics how humans universally communicate intents with each other using language. However, there is much more than language that leads to effective human communication [43]. Further, humans' expression and understanding of each other's intents can be inherently ambiguous and often not perfect—how can we expect to do it better with Generative AI?

One obvious solution for improvement is providing better media for intent expression, such as combining multiple modalities: text prompt, sketch, and gesture. However, the ambiguity of intent could be much more fundamental in that the human might not really know what they want (Generative AI) to create in the first place. As such, it might be useful to maintain a continuous conversational between users and Generative AI, rather than expecting to arrive at ideal results in one-shot attempts. Some intents are implicit, assumed by the human but often unspoken. For example, a user generating a furniture with a computer-aided design (CAD) tool (*e.g.*, [38, 120]) might look good on the screen yet they also implicitly expect it to look equally good when manufactured and placed in the intended environment without realizing that the Generative AI does not know about the manufacturing process or what the environment is like.

To address the inherent challenge of ambiguity in intent expression, our next-steps for HGAI span both input and output of Generative AI (Figure 3) with an emphasis on rethinking the dominant use of text prompts.

## 4.1 Explainable and Guided Text Prompting

Given the overwhelming popularity of text prompting in numerous Generative AI scenarios, some future efforts on HGAI should be devoted to better supporting such input with explanation and guidance, which can be helpful for both end-users and developers.

*4.1.1 Enabling end-users to understand and manipulate input/output relationship.* End-users often do not know how good is the text prompt they use in getting Generative AI to produce the result they are looking for. Similar to how adversarial examples lead to unexpected errors in Discriminative AI [63], the analogous issue exists in Generative AI when the changes a user makes in the text prompt fails to produce the changes they expect to see in the generated contents or, worse, produces undesirable changes. The opaque or unexplainable relationship between text prompt input and generated output often leads end-users to run unguided trials that drains their time and wastes computing resources. As a result, users might have a hard time establishing trust and willingness to accepted generated contents [27]. Some next-steps for HGAI include the following.

More studies should be conducted to understand how humans use prompts to interact with Generative AI, such as when writing text [46] or conversing with a chatbot [151]. Such studies should aim to provide concrete evidence that complements the currently anecdotal understanding of prompting and to further connect with Generative AI researchers and developers to inform their model-building work.

Generative AI systems should provide tutorials and examples that educate end-users about the non-deterministic behavior of Generative AI and manage their expectation, thus maintaining a reasonable level of user's trust in the model while preventing unguided prompt engineering.

Developing feedforward techniques [18] to visualize what certain edits in a text prompt might lead to changes in the generated contents. Although there have been studies on the effects and trade-offs of such feedforward controls (using pre-generated examples) [47], it remains unclear how to implement such techniques at interactive speed without requiring pre-generated examples.

Similar to feedforward, borrowing the autocompletion approach in text entry [37], we can develop techniques to suggest words or phrases following a user's partial text prompt and optionally show what contents will be generated if provided with the completed text prompt. Importantly, the user should be able to scroll through multiple autocompletion candidates to explore which one best fits their intent.

For explaining Discriminative AI, past work has employed attention models and saliency maps (*e.g.*, [130]) to indicate which parts of the input is "responsible" for certain output. Analogously, we can develop and integrate similar approaches for Generative AI, *e.g.*, the Cross Attention approach [135], allowing the user to explore and understand how their text prompt is associated with the generated output. Further, we can develop techniques for the user to directly manipulate the output (*e.g.*, certain parts of a generated image) and see, inversely, how the input text prompt changes.

Enabling controlled generation (discussed later in this section) can also contribute to the user's understanding as it allows a user to manipulate specific generation-controlling modules and see its causal effect on the generated results.

*4.1.2 Enabling developers to discover and troubleshoot problematic input/output relationships.* Beyond obtaining a satisfactory result, it is also important for a Generative AI model to respond appropriately to certain changes in input. Similar to the general concept of sensitivity analysis, some recent work has found suboptimal or problematic input-output response relationship in Generative AI, *e.g.*, the orders in which training samples are provided result in drastically different performance [96] and over half the tokens in a prompt can be removed while maintaining or even improving model performance [149]. Building off of these findings, one next-step for HGAI is developing tool support for sensitivity analyses of Generative AI that enables model or application developers to surface problematic responses that might otherwise go unnoticed. Such tools can provide a useful dashboard that monitors and visualizes model responses given a set of benchmark input perturbation.

## 4.2 Restructuring the Workflow of Text Prompts

Going beyond the monolithic "prompt-revise-repeat" cycle, HGAI research should propose alternate workflows that allow a user to break down a complex generative task or to efficiently iterate on suboptimal generated contents. Here we focus on discussing one specific alternate workflow–the insertion-based control approach.

*4.2.1 Iterative insertion-based control of Generative AI.* Rather than a one-shot text prompt input, a new workflow can start with something simple, which then allows for iterative insertion of additional input to extend or refine generated contents. InsNet is one technical solution for such insertion-based control where the method generates sentences in random orders by inserting tokens to existing partial contexts [95]. Another approach employs smaller models to control larger models where users can, for example, stipulate that they want three specific words or a specific style to appear in the generated text [48, 98, 153]. Next-steps for HGAI along this direction include the following.

Building off of recent work on chaining prompts [147] and models [52], we can develop more varieties of mixed-initiative workflow for natural language generation. For example, a user writing a story can start with a single word, *e.g.*, "flower". Next, the system prompts the user to describe how they feel about flowers, *e.g.*, "I love flowers", which then allows the natural language generation methods to insert new elements to create longer and more sentences. At each turn, the user can also insert their own elements before AI takes over. In the meantime, the user interface displays the history of iterations showing how the text "evolves", thus allowing the user to roll back if the expansion has taken an undesirable direction.

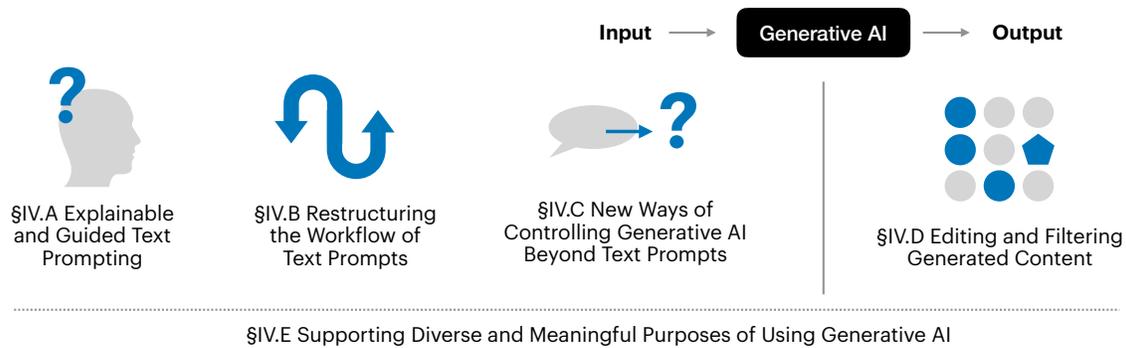

Figure 3: Overview of HGAI Level 2: next-steps in assimilating human's expression of intents.

In the image generation domain, we can develop a workflow where a user starts with a simple text prompt and Generative AI returns an image. Next, the system generates a textual description based on the image and extract words associated with visual elements of the image. The user can then edit the generated image by manipulating the corresponding words in the textual description. The LUCSS demo[8] showcases a proof-of-concept prototype related to this approach, where a user can colorize an initially black-and-white generated image via manipulating color attributes in the textual description [158].

## 4.3 New Ways of Controlling Generative AI Beyond Text Prompts

Currently, the interaction design of prompting mainly assumes that humans initiate and AI reacts, while ignoring other types of possibilities (*e.g.*, human-machine co-creativity and mixed initiatives), as theoretically constructed as a $2 \times 2 \times 2$ design space in [91]. Although text prompts leverage humans' familiarity with using language to express intents, in the meantime, it also limits other forms of expression that also exist in human-human communication, such as visual language, gestures, and facial expression. For example, research in psychology [97] has found that gesture is more than an auxiliary aid to speech but rather an integral part of human communication and that it is closely linked to thought, working together with speech to develop meanings. An even more fundamental issue is, regardless of what forms of expressions are available, users themselves might not always know their intent (*i.e.*, what exactly they want to create). It is possible that a user's intent evolves and clarifies itself as they iteratively attempt to express it to Generative AI and to revise the input based on the generated contents. Current text prompting interfaces have not been specifically designed to support such intent exploration—a successfully "engineered" prompt often looks remotely like humans' natural language expression and is challenging for non-expert users to come up with [151].

---

[8]LUCSS: Language-based User-customized Colorization of Scene Sketches (LUCSS): https://youtu.be/IsBdrXtU0MI

*4.3.1 Spatial and gestural input to control Generative AI.* In contrast to the "1D" text prompt, input to Generative AI can leverage an additional degree of freedom and we should explore how 2D or 3D techniques can allow users to express their intents in various domains of creation. Specific next-steps for HGAI are as follows.

Existing techniques, *e.g.*, ControlNet [154] allows a user to condition image generation with additional images, *e.g.*, one with edge detection to provide a skeletal "template" for the model to fill in generated details. Building on and going beyond this approach, we can study what kinds of templates a user would create to express their intents to the model, which will likely lead to new features currently unsupported by Generative AI models. For example, in the medical domain, a pathologist specifying the generation of synthetic histopathological images of tumor cells might provide very different kinds of sketches than an architect outlining a new concept of office buildings.

Expanding input to 3D, we can expect crosspollination with gestures and augmented or virtual reality (AR/VR). Beyond using generated contents to construct the AR/VR world [72], another challenge and opportunity is enabling users to create generated 3D objects (*e.g.*, using Shape-E [12]) such as furniture or art installations. For furniture design, one direction is to integrate some existing techniques (*e.g.*, freehand gestures [70] and 2D+3D sketching [15]) with text prompting and the latest Generative AI models.

Even in the natural language domain, *e.g.*, writing, prior work has demonstrated the possibility of using a brush-like input to directly manipulate key attributes of the story, such as the fortune of the protagonist character [41]. One next-step is to explore how users can control other parameters in text generation using 2D input techniques, including the usage of a dashboard that presents comprehensive key parameters of generated text using data visualization techniques [102].

*4.3.2 Controlling Generative AI with implicit intents.* Some intents are implicit and naturally unspoken, assumed to be understood by others where appropriate actions should take place accordingly. Examples include contexts, activities, and personal history. Building on recent developments such as zero-shot multimodal reasoning [152], next-steps in HGAI should aim to incorporate these implicit intents:

Leveraging a large body of work on context-aware computing [121] and activity recognition [36], we can incorporate additional information as representations of a user's intents. Such an approach can be useful when employing Generative AI to automate physical tasks (*e.g.*, via controlling a robot or an Internet-of-Things). For example, as a user wakes up and walks towards the kitchen, the time and location contexts can inform the Generative AI model to reason that the user might want to make coffee and proactively generate next-step actions to start the coffee machine.

Some user input or edits might represent implicit intents that Generative AI needs to incorporate when creating certain contents. For example, consider a furniture designer using a CAD tool to put together the tabletop and three legs, which carries the implicit intent that these legs need to maintain contact with the tabletop. As such, Generative AI that morphs the shape of the tabletop should also reposition the legs to maintain contact.

Personal history (*e.g.*, daily routines) can also be helpful, such as in the above automating coffee-making example. Although generative conversational agents like ChatGPT or Bing Chat do retain certain history, much more work needs to focus on how domain users utilize history in their work, such as programmers debugging a large codebase or doctors examining a patient, which should, in turn, inform the development of new Generative AI models that can leverage such historical data to improve generated contents, whether it is a code snippet or a summary of patient history.

## 4.4 Editing and Filtering Generated Content

Whereas the above discussion in this section focuses on reinventing the input techniques, below we switch gears to consider end-user editing and filtering of Generative AI's output. When a user is unsatisfied with the generated contents, rather than restarting the whole generation process to obtain a new sample, it would be more efficient for the user to specify what is not right and directly edit the generated contents.

*4.4.1 Semantic control of generated contents.* Prior work has demonstrated "smart" edits where a user's drawing or erasing of the generated design translates into new constraints that steer the AI to generate a new version while addressing the user's intent [38]. A series of similar approaches have emerged when working with GANs, from providing sliders to adjust various attributes factorized from the latent space such as pose and texture [125] to allowing for directly dragging the generated image [111, 141]. Building upon these methods, some next-steps for HGAI are as follows.

It is important to provide such semantic controls with continuous feedback at interactive speed, which might require new interpolative and approximative techniques beyond just generating whole brand-new contents every time a user dials the knob. We can conduct studies and analyze the "knob-dialing" behavior of end-users when provided with semantic controls and understand what is the minimum amount of feedforward information we can provide to inform users' control without adding latency to the system.

A more fundamental approach is to change what AI generates: not pixels or tokens but semantic controls, which can offer users not just a static result but also a tool to access a large space of alternative contents. Recent work that simplifies an image into highly abstract yet representative sketches [140] shows promises of providing such semantic controls as the outcome of the generative process. Relatedly, Videomap lets a user perform video editing by navigating on a 2D view of the latent space [90]. Generating controls beyond texts or pixels requires both technical breakthroughs as well as studies to verify that such controls would allow a user to realize their intents without too much cognitive load.

Given how generated contents are likely to contain imperfect or flawed elements (*e.g.*, blurry faces in generated images), we should develop techniques for users to fix such issues as directly and quickly as possible. Prior work has demonstrated some user-driven steps for fixing entanglement issues in GAN [55], such as specifying regions on the generated images that should be disentangled. Similarly, we can employ other techniques to support direct fixes, such as using image inpainting techniques [150] to redraw blurry faces.

In the natural language domain, we can explore techniques to present generated output (*e.g.*, a chatbot's response) in a more editable way to collect users' immediate feedback that informs the next iteration. Today's chatbots are not yet operating at the level of a competent human collaborator. Instead, they are more similar to primitive APIs that are accessible via chat interface. The lack of intent understanding and context awareness make them hard to control and iterate.

As a workaround, chat experience developers can augment text-only chat with additional controls. For example, consider summarizing a news article. The output summary can come with "+/-" buttons to adjust its length or allows a user to highlight portion of the text as "important/unimportant" so the model can create a new summary accordingly. A similar approach has been demonstrated in a text reader where a user's highlights in an article can steer the summary to put more emphasis on the highlighted texts [39]. In the longer-term, we envision the capability of a chatbot will evolve to serve better as a collaborator instead of a passive API.

*4.4.2 Filtering generated contents.* Alternative to directly editing the generated content, a user can express their intents via selecting which types of results they prefer over the others, which sends a signal to the model for generating more relevant contents in future iterations. Such approaches have demonstrated expressive power when the user is faced with a large number of data points, such as document collection [44] and GAN editing directions [54]. Specific next-steps for HGAI include the following.

Developing methods to represent and extract user intents from the selection they perform, which the Generative AI model then incorporates as a signal into the subsequent generation of new contents.

Conducting studies of such a user-driven coarse-to-fine selection of generated contents to compare key metrics (*e.g.*, content quality, cognitive load, user satisfaction) with conventional approaches that involve repeated generation with prompt tweaking.

Assessing the potential risk of generating a wide range of unexpected content for users to choose from, some of which might fall way out of distribution and only serve as noises and some might even include inappropriate elements such as toxic language or disinformation.

## 4.5 Supporting Diverse and Meaningful Purposes of Using Generative AI

Finally, human intents of using Generative AI would also be influenced by what Generative AI is capable of. In contrast to AI-generated contents as commodity (*e.g.*, generating an attention-grabbing video optimized for virality rather than deep meaning), one next-step for HGAI is exploring designs that support more diverse and meaningful reasons for humans to use AI-generated contents (*e.g.*, generating a video to tell the life story of a family member). Consider using Generative AI for communication, such as parents telling bedtime stories to children. To support such communicative purposes, the focus of Generative AI should go beyond producing a stereotypical story and aim to support conveying cultural meanings, cultivating familial relationship, or even allowing parents to teach children a specific lesson metaphorically via storytelling.

## 5 HGAI NEXT STEPS: AUGMENTING HUMAN ABILITIES

Generative AI is more than a computational model or a tool; it will significantly change how professionals are able to work. For example, filmmakers might use Generative AI in the video editor to create a shot they forget to take. With such a capability, filmmakers no longer need to take footages exhaustively and worry about missing some shots. Further, Generative AI will change how people perceive their profession just like how algorithmic automation has been changing work in many fields [132]. For example, an illustrator might rethink what it means to create a piece of work given how Generative AI can automate partially or even mostly what they do. Generative AI will blur the boundaries between professions. One example is the blending of art and engineering: people who are familiar with the inner-working of Generative AI and good at "prompt engineering" can explore the creation of artworks whereas artists can tap into the engineering world to harness the power of industry-scale models via fine-tuning.

This section's discussion of HGAI next-steps (Figure 4) covers domain-specific data and content generation, teaching domain knowledge to Generative AI, and integrating Generative AI into domain users' workflow. Finally, as a research community, we also started to wonder how Generative AI can augment (or otherwise affect) our abilities to conduct research, which we further discuss in the last part in this section.

## 5.1 Generating Domain-Specific Data and Contents

One important way for Generative AI to augment domain users' abilities is to help them overcome the hurdles of acquiring data, which could be costly and time-consuming due to data scarcity and limited resources.

*5.1.1 Generating 3D objects and scenes.* Although there exist solutions to generate 3D objects and scenes (*e.g.*, [77, 146]), there remain gaps in making such generation useful for specific domain users. We discuss a few exemplar cases below and their next-steps for HGAI.

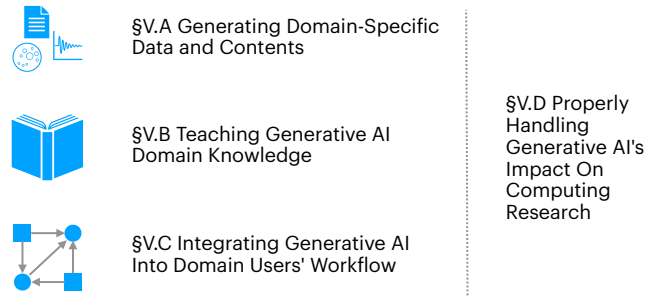

**Figure 4: Overview of HGAI Level 3: next-steps in augmenting humans' abilities in a collaborative workflow.**

Consider generating traffic data for training self-driving systems. Current Generative AI cannot generate safety-critical traffic scenarios because the model learns to fit a data distribution and sampling from such a model can result in the most probable synthetic sample with limited training value. In many applications, self-driving systems care about corner cases or low-tail samples, *e.g.*, accident-prone traffic scenarios. One important next-step is to enable Generative AI to "extrapolate" and synthesize less frequent but safety-critical scenarios. The same concept can benefit LLMs and text-to-image models to provide even more creative output, instead of giving an average answer based on its large pool of training data.

Consider generating 3D objects for digital design and fabrication. Currently, Generative AI can only create static 3D models (*i.e.*, point clouds or meshes) that cannot be easily modified to fit domain users' different needs. One next-step is to generate machine code (*e.g.*, G-code) that drives a fabrication machine to create an object so that domain users can modify the code for custom designs. One such example is generating 3D printed hair [85]: rather than generating the geometry of hair, it is more customizable to generate G-code where a user can directly manipulate key parameters, such as length, thickness, and curliness.

Consider generating architectural designs such as floorplan layout and furniture arrangement. Beyond the current approaches that focus on generating static plans [60, 106], one next-step is also generating a simulation [112] of how people interact with each other within the space to better inform architects and designers to further iterate on their work.

*5.1.2 Generating medical data.* Data has always been both the fuel and the bottleneck of medical AI development due to the data scarcity of certain diseases as well as the high cost of collection, processing, and labeling. Multiple opportunities and challenges exist for the future for HGAI.

Given the recent development of synthetic medical data generation (*e.g.*, in histopathology [51] and radiology [32]), it is time to study the effects of using such synthetic data in downstream tasks, especially on medical AI models' performance on out-of-distribution datasets as well as how doctors and patients perceive such models knowing the training data is "not real". Further, to address accountability issues, it is important to allow doctors to trace an AI error to potentially problematic synthetic data and verify its factual correctness [104].

Besides medical data of different pathologies, it is also possible to use Generative AI to create "synthetic" control patients in clinical trials [6]. One related challenge is providing tool support for experimenters to carefully control Generative AI when manipulating parameters of "synthetic" control patients and to generate reports with transparency to fully inform policymakers the limitations and potential risks.

However, we should also be on alert of the dark pattern related to this direction: generated, faked clinical trail data in medical research and publications, which has been a long-standing issue that can be further exacerbated by advances in Generative AI [138].

*5.1.3 Generating contents across traditional boundaries of formats.* For art and design, one opportunity and need is to support media objects and systems that can naturally cross traditional boundaries (physical/digital, 2D/3D, interactive/static, raster/vector, authored/generated). Specific next-steps for HGAI are as follows.

As mentioned in the previous section, image generation should consider providing users with editable contents beyond static pixels, *e.g.*, vector graphics or design tool files (*e.g.*, .ps and .ai). One additional benefit of this approach is that users can perform edits on these files, which offers process-oriented information to instruct AI how to generate next-iteration contents to better meet the user's needs.

An even further goal is for AI to define and generate an "invariant representation" of an object that can be malleably converted into a wide range of formats so that users can readily import the generated contents into their work using different tools. Consider multimedia industry creating a new character that spans comic books, animations, movies, theme parks, and video games. Generative AI will create the "core" of the character, which is then expanded to different media. Any future updates or additions to the character will also be automatically and consistently reflected in individual types of media. The benefits of this approach are that *(i)* Generative AI can obtain and take in feedback from heterogenous types of users (*e.g.*, comic book readers, movie viewers, and theme park goers) and that *(ii)* Generative AI can serves as the nexus connecting different departments within a company or industry to co-develop the character.

## 5.2 Teaching Generative AI Domain Knowledge

Current Generative AI models indiscriminately scrap and learn from data on the Internet without explicit recognition of domain knowledge. As a result, some types of generative contents are problematic. While synthesizing texts from a book without knowing the subject matter might still produce sensible writing [107], synthesizing pixels from images of human hands does not work as well due to the ignorance of anatomy [34]. There are numerous other cases where Generative AI's lack of domain knowledge will cause performance issues, *e.g.*, generating drug designs, protein structures, molecular models, building codes, and industrial manufacturing equipment.

One popular approach to overcome the knowledge gap is in-context learning [101] where an end-user provides an example (*e.g.*, an existing story) and asks Generative AI to produce something similar (*e.g.*, a new story of the same genre). However, this approach likely will not work across many other domains, *e.g.*, generating a W2 tax form that requires much more knowledge available in a single example. Alternatively, retrievable augmented generative models [155] can adapt to highly specialized domains. We can still train general-purpose large models but have small, domain-specific knowledge bases to retrieve from and use the retrieved results to augment the black-box general language models to quickly adapt to new domains.

Below we discuss next-steps to teach Generative AI two types of domain knowledge: "what" and "how".

*5.2.1 Teaching Generative AI "what-knowledge" by concepts.* One common way to represent knowledge is hierarchical concepts, such as the description and organization of medical conditions [23]. Concepts can be thought of as "what-knowledge" as it informs us what makes an object what it is and different from the others, *e.g.*, defining a human hand by its constituent parts as well as their spatial relationship. Currently, Generative AI is oblivious of concepts, *e.g.*, able to generate an image of a bike but unaware of the correspondence between different parts of the image to specific parts of the bike. Research on Discriminative AI has realized the importance of teaching concepts to a model [28, 35, 79, 84], mainly for interpretability purposes and to ensure that AI is right for the right reason [117]. Meanwhile, concept teaching in Generative AI remains a nascent research topic with some next-steps as follows.

Similar to how Concept Activation Vector [79] uses positive and negative examples to represent a concept, we can develop Generative AI models that can follow user-specified concepts. GANravel is a tool that employs this approach, letting users select example images to unbias GAN's image generation [55]. Examples can more effectively and intuitively represent a user's domain knowledge where textual expression falls short (*e.g.*, due to inherent vagueness or under-defined terms), such as pathologists describing pathognomonics—visual features of certain types of tumor cell.

Rather than having multiple models learning different modalities of data, we should develop Generative AI that learns symbolic concepts (*e.g.*, door), which can be represented equivalently in different types of generative contents (*e.g.*, an abstract icon, a photo-realistic image, the sound of a door opening, mechanical behavior of door knob and hinges). In this way, Generative AI can produce comprehensive contents necessary for the user's task, such as a video showing a person opening a door. To achieve this, one challenge is the need for pairs of data (contents and the constituent concepts) and one solution would be using CLIP [114] to construct a concept dataset from images to texts.

The recent development on image segmentation [80] shows promises in "dissecting" static images into components, which can support concept learning. However, we should aim further to establish hierarchical relationships between parts. In so doing, the model can learn if different parts are functionally similar, *e.g.*, scissors and a knife both have blades. Building on the "library learning" approach in program synthesis (*i.e.*, discovering library components or subroutines in a program with semantic meaning) [53, 145], we can explore Generative AI that learns to generate instructions of creating certain contents (*e.g.*, G-code) and to extract concepts represented by components within such instructions.

*5.2.2 Teaching Generative AI "how-knowledge" by examples and demonstration.* Complementary to concepts as "what-knowledge",

teaching Generative AI how to create certain contents serves as "how-knowledge" that can bridge the gap between a generic model and users' domain-specific needs. Some next-steps for HGAI are as follows.

Taking the programming-by-demonstration approach, we can allow domain experts to perform a creative task for Generative AI to learn and imitate. In the Discriminative AI domain, past work has demonstrated users' teaching an object recognizer in real-time [156]. For Generative AI, for example, an artist drawing caricature in some unique styles can demonstrate the key steps they follow; Generative AI, in turn, can learn to perform these steps, each of which would allow the artist to tweak, adjust, or innovate in their familiar ways. The grounded generation approach [88] provides a nice starting point that can allow artists to work on semantically separated elements in the generated image either by adjusting the prompt or by direct edits.

On the other hand, some *a priori* domain knowledge should be incorporated during model training, rather than having to be demonstrated by each user when interacting with the model. For example, for scene generation, the model should ideally learn to provide camera controls (*e.g.*, aperture, focal length) as they are well-known parameters a human photographer or cinematographer would want to control.

## 5.3 Integrating Generative AI Into Domain Users' Workflow

Perhaps the greatest challenge and opportunity in this HGAI level is integrating Generative AI appropriately into a domain user's workflow based on a solid understanding of how they work, what is the best role for Generative AI, how to augment the user along each step, and what task-related contexts can further inform Generative AI.

*5.3.1 Understanding humans' mental model of collaborating with Generative AI.* In terms of mental model, is collaborating with Generative AI similar in some way to collaborating with other types of computing systems or collaborating with humans? As suggested by prior work on chatbot [78], a more fundamental understanding of Generative AI users' mental model can influence user experiences and inform design decisions: whether we should fit Generative AI in the old ways of work or it is worth defining a new eco-system of work unique to Generative AI. One lesson from an analogous domain is the use of freehand gestures. When camera-based tracking became widely available (*e.g.*, Microsoft Kinect), some applications simply mapped freehand gestures to GUI buttons (old ways) rather than exploring more natural and expressive input superior to button pushing [143]. Next-steps for HGAI should carefully consider users' mental model when employing Generative AI, such as in code generation as an example. For programmers, simply integrating prompt-based code generation into their IDEs might be insufficient to fully realize Generative AI's potential in augmenting their programming abilities. More nuanced designs should consider a broad spectrum of issues: how long the generated code should be, when to trigger single *vs.* multi-line code, whether to provide single or multiple suggestions, how much latency is acceptable, which information to condition the model on (all files open in the IDE *vs.* the single file in focus), how to communicate to the programmer what information is being "read" by the model, how to allow for a model not to have access to sensitive files, and how to onboard users to learn all functionalities of the tool.

Based on a well-understood users' mental model, one central question to answer is where to find the best place to use Generative AI in a user's work, which we discuss next.

*5.3.2 Finding the right places for Generative AI.* Although Generative AI promises to provide on-demand contents throughout a user's workflow, it remains unclear where a user should employ Generative AI, how, and how much is the utility of incorporating generated contents compared to conventional approaches. For example, image generation sounds useful for visual designers but sometimes retrieving contents from stock images (*e.g.*, a photo of a McDonald's restaurant) is already fairy convenient and will have no quality issues that some Generative AI models suffer from at times. For such tasks, Generative AI can replace but will do no better than conventional approaches. On the other hand, if the contents needed cannot be easily found (*i.e.*, out of distribution, such as an underwater McDonald's restaurant), then Generative AI will play an indispensable role and save much efforts (*e.g.*, searching for and Photoshoping the non-existing image). To find the right places for Generative AI, some next-steps are:

Prior work on human-AI collaboration surfaced two "camps" of involving AI in human's work: a top-down approach where AI reports findings to a domain user (*e.g.*, via a hierarchical organization of diagnostic evidence [64]) and a bottom-up approach where AI acts as a "copilot" to assist individual steps performed by a human (*e.g.*, recommending where to examine on a medical image [65]). Analogously, we can instrument Generative AI either in a top-down or bottom-up manner to assist creators' work and study their reaction and preference between these two canonical approaches.

Meanwhile, it is also worth studying unique issues of Generative AI—the gap between what a model is expected to do to help domain users and what domain users actually find useful for their work. For example, to find out whether LLMs can help screenwriters with their scripts, we need to understand how screenwriters go through different stages of writing scripts and then identify at which stages they are most likely to use LLMs and how. One recent project has studied and developed tools for 3D designers to use text-to-image generation in their workflow [93]. Importantly, surfacing unexpected usages as well as non-usages will inform the development of next-generation Generative AI models and the eco-system of tools.

There is the expectation that Generative AI should magically act as the genie that grants users' wishes of specific contents they are unable to create on their own. However, for creators that pursue original works, another valuable use of Generative AI is to support early-stage exploration. Thus creativity support tools powered by Generative AI should focus the interaction design on encouraging back-and-forth iterations, presenting diverse somewhat-optimal contents (rather than narrowly-defined optimal ones), and tutorials for making things based on which the creator can extrapolate and expand on their own.

Collecting training data that not only includes the final outcome of a creator's work but also intermediate data that documents the process, *e.g.*, different versions of a drawing from rough outlines to sketches and to a version with fine details. Such a dataset would allow us to benchmark Generative AI's performance at different steps and recommend when a user should involve AI. Collectively, datasets like this across various domains provide evidence for developing a theory of what human and Generative AI are good at, respectively.

*5.3.3 Step-by-step generation grounded on specific domain knowledge.* Although aiming at the same final goal, human and AI might take very different approaches. In early research of medical AI, Blois found that human doctors' diagnosis often follows a funnel-like process [22]: starting with broad hypotheses, then running tests to gradually narrow down possibilities, and finally confirming the most probable cause of the observed symptoms. In contrast, most medical AI models only did best towards the end of the funnel (*i.e.*, telling whether the patient has disease X) but not so well at the triage step at the beginning.

Similarly, in most human creative process, whether it is writing a story or painting a picture, the creators would develop a domain-specific step-by-step approach, which is rarely reflected in Generative AI that achieves the same content creation. Generative AI is generally unaware of any intermediate steps and only aims for the tokens or pixels in the final result. Although it is possible to use Generative AI to simulate the step-by-step approach, *i.e.*, generating intermediate artifacts and using them as input for further generation. There is no guarantee that the result will be superior to the one-shot approach. Some next-steps for HGAI are as follows.

Studying whether and how end-users simply utilize Generative AI to obtain the final result or there exist attempts to perform a step-by-step workflow by generating intermediate results to build on. A systematic study (*e.g.*, using technology probe) can surface a design space of using Generative AI throughout the entire creative process.

Focused on specific domains, *e.g.*, architects designing buildings, conducting studies to understand human creators' step-by-step workflow in their existing practices, based on which we can assess whether a Generative AI module can support each step.

Evaluating the one-shot *vs.* step-by-step approaches, comparing the qualities of generated contents as well as end-users' agency, workload, and satisfaction. One hypothesis is that one-shot generation is better for the initial exploration step whereas step-by-step generation is better when the creator has identified a specific direction and wants to incorporate their own creative elements into the generated contents.

*5.3.4 Imbuing Generative AI with task-related contexts.* As we are expected to invoke Generative AI frequently throughout our workflow, it is important that Generative AI should obtain as much contextual information as possible. To achieve this, some next-steps for HGAI include the following.

Study what task-related contexts domain users make use of in their work and whether such information can be used by Generative AI. For example, for creators in theatre, given a spatial configuration of speakers, Generative AI can suggest optimal acoustic effects for best experiences; for artists exploring multiple displays, Generative AI can propose uncommon sequences of visuals. Another example is involving intelligent tool support (not Generative AI per se) in designers' ideation process where they would draw on materials to construct a mood board [83]. A current Generative AI model might assume designers can change their process and use text prompting to help their ideation; yet the above project demonstrates how embedding support into their familiar workflow ("designer-led") based on task-related contexts is a more pragmatic approach.

To obtain relevant task-related contexts, one promising approach is integrating Generative AI with AR systems equipped with sensors that can recognize real world objects and scenes. Imagine a user asking Generative AI to come up with a recipe based on what ingredients they have. Integrated with AR and sensors, the system can map each generated step onto specific ingredients and track the cooking progress, which is a much more immersive and natural experience than just relying on generated text recipes.

Another open challenge and opportunity is how Generative AI can help multiple creators collaborate (*e.g.*, human-human ideation [128]), which requires an understanding of creators' dialog with each other and what contents AI should generate that can catalyze collaboration. One analogous project focused on human-human communication where the system retrieves relevant images by inferring when a user might find it useful to have such visual information to illustrate their speech in a video conference with others [94].

## 5.4 Properly Handling Generative AI's Impact On Computing Research

At the end of this section, we dedicate some discussion to how we can augment computing researchers' abilities by properly handling Generative AI's impact on how research is conducted.

*5.4.1 Handling changes in conducting HCI research due to Generative AI.* Beyond the obvious (and non-HCI-exclusive) use of LLM to assist writing, there are other emerging changes in how we conduct HCI research that need to be handled properly as next-steps.

Some qualitative coding software already introduced LLM-based code generation (cf. a discussion [50] on LLM for thematic analyses [24]), which trades off original interpretation with convenience. Qualitative research is subjective, as one project their identity into the interpretation of the data; using AI loses one's position and subjectivity. The HCI community should develop guidelines, reporting requirements, and reviewing criteria of research that analyzes qualitative data using LLMs.

It is not the first time that computing research has to grapple with redefining research contributions in light of new advanced techniques. When deep learning first became a popular tool, gesture recognition algorithms found a hard time to claim a contribution given how deep learning models could already achieve a higher performance than many handcrafted algorithms. Learning from this historical lesson, as a next-step, technical HCI research [? ] should define new agenda, focusing on inventing new interactive systems that catalyze Generative AI to support users to achieve more than what a vanilla Generative AI model can offer.

On the other hand, using mature and high-performing off-the-shelf tools (be it deep learning or Generative AI) might soon be disregarded as contribution. To handle such changes, the HCI community need to renew our definition of what constitutes an artifact contribution [144], addressing possible reviewer questions like "when is using LLM in building a system considered a contribution?"

While a plethora of HCI research will soon flourish by building useful tools based on Generative AI, the community should ensure equal, if not more, emphasis on tools that prevent or mitigate harm done by Generative AI, from preventing programmers from over-relying on LLM-based code generation [58] to reducing the chance of training models on artists' work (*e.g.*, by adding adversarial noises [123]).

The HCI community can promote tool building that supports researchers to properly and productively use Generative AI, such as following Soylent's approach [20] to let LLM automate the tedious and non-intellectual parts of paper writing. Another direction is enabling junior researchers to exchange ideas with Generative AI and obtain quick feedback to their work.

*5.4.2 Cross-disciplinary collaboration (HCI + Generative X) will become a necessity.* Consider the emergence of LLMs, most notably OpenAI's ChatGPT, which has democratized Generative AI to the vast public where models are no longer evaluated by benchmarks but directly judged by end-users. Therefore, there is an opportunity to introduce human subject evaluation methods to NLP research. On the other hand, HCI research is no longer constrained by a lack of NLP expertise because industry-scale models are now easily available and can be fine-tuned to fit specific use cases. Given how NLP and HCI develop the need for each other, one natural next-step is promoting the norm of collaboration across the two fields and further across HCI and other "Generative X" domains.

## 6 CONCLUSION

Currently, Generative AI is one of the fastest growing fields; yet, we argue that focusing too much or pursuing some immediate research ideas might lose sight of a holistic picture that can connect multiple disciplines towards some long-term, shared missions. Unifying a wide range of ongoing topics as well as less-explored ideas, this paper contributes a research agenda that lays out the landscape of next-steps for HGAI. Specifically,

- We define the term "Human-centered Generative AI" (HGAI) as three levels of objectives: aligning with human values, assimilating human intents, and augmenting human abilities.
- We followed a structured process to iteratively formulate next steps for HGAI across all three levels into a coherent and comprehensive research agenda.
- Our proposed next steps cross disciplinary boundaries and draw on insights from both academic and industrial research, thus inviting members of multiple communities to collectively tackle future challenges in HGAI.

One current limitation is a lack of direct involvement with diverse stakeholders who are impacted by Generative AI (*e.g.*, designers and artists). Mitigating harm caused by Generative AI itself is another grand challenge, as indicated in this comprehensive report [9] by epic.org. We plan to address specific stakeholders in the future as we pursue specific topics identified in the proposed agenda.

We hope these next-steps can serve as starting points for researchers across disciplines to collaborate and pursue specific ideas while staying informed of the big picture. As Generative AI continues to develop at unprecedented speed and scale, we believe that taking a human-centered approach early on can have a significant long-term impact on the future of human-AI symbiosis [89].

# APPENDIX: PROCESS OF DEVELOPING HGAI NEXT-STEPS

To formulate the research agenda for HGAI, we conducted three iterations of discussions over a period of 1.5 months. Our participants ($N = 11$, two female and nine male, aged 28-46), *i.e.*, all the authors, include five academic and six industrial researchers whose research expertise span technical HCI research, machine learning, natural language processing, computer vision, and computer graphics. All participants had prior experiences developing or employing various generative methods in their research (Figure 5).

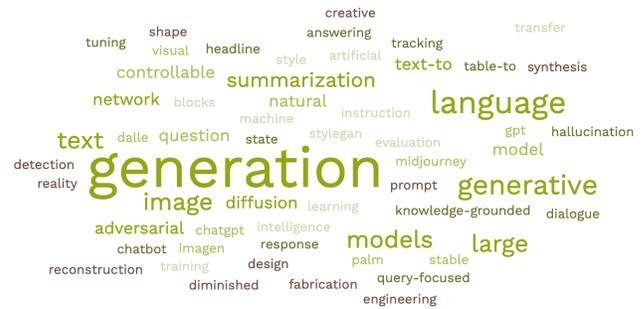

**Figure 5: A word cloud visualization to illustrate the authors' expertise and experiences related to Generative AI based on their own provided keywords.**

## 6.1 Iteration #1: Individual Brainstorming Discussion

We started with individual brainstorming discussions between the first author and each participant, *i.e.*, 10 rounds of 1:1 conversations. The main purpose was to establish the breadth of an HGAI research agenda by generating as many ideas as possible. In each discussion, we spent the first five to ten minutes coming up with possible ways of defining HGAI, following which we brainstormed HGAI-related problems and new topics for future research. The participant was the primary contributor and followed the think-aloud protocol [86] while the first author acted as an interviewer and note-taker who focused on prompting the participant to elaborate, clarify, and broaden their ideas. We intentionally avoided delving into each idea as we focused on breadth in this iteration while leaving deeper discussions later. Each 1:1 discussion lasted between 45 minutes to an hour: all but one discussion was conducted in-person.

After the discussions, the first author summarized the notes of each discussion into a list of research agenda topics with brief descriptions. Further, we aggregated participants' proposed definitions of HGAI and divided them into three levels of interpretations as detailed later in §2.

## 6.2 Iteration #2: Paired Discussion

Next, we conducted five discussions that each involved the first author as the moderator and two other participants. In all but one discussion, the two participants were split between academia and industry. In all discussions, the two participants had different areas of expertise. The main purpose was to identify interdisciplinary HGAI research opportunities, such as common problems shared by multiple disciplines and innovative system designs by combining multiple disciplinary elements. Each discussion revolved around three research agenda topics selected based on the previous discussions: we first selected topics both mentioned by the two participants, which resulted in either one or two topics; then we selected the remaining one or two topics amongst the ones with the most extensive discussions in the previous iteration (measured by the amount of notes). For each topic, after briefly describing what was discussed before, we asked the participants to think further about prior work related to the topic, the gap in said topic, and specific research activities to pursue the topic. Each discussion was conducted remotely and lasted for about 45 minutes.

After the discussions, the first author followed the Affinity Diagram approach to organize notes from the previous two discussions into tree-like structures: each research agenda topic was the mid-level node, whose low-level nodes consisted of prior work or specific research activities for future work, and each top-level node was a theme to connect multiple related topics.

### 6.3 Iteration #3: Virtual Walk-the-Wall Discussion

Finally, we asked each participant to walk through the Affinity Diagram laid out on a shared document. Participants could add to the low-level nodes, suggest or edit research agenda topics at the mid-level, or add or re-organize the top-level themes. We tracked the changes made by each participant and the first author facilitated ad hoc discussions whenever there were conflicted changes or disagreements amongst participants. This walk-the-wall discussion took place asynchronously over a period of 15 days, after which the first author finalized the additions and changes to create a clean version of the Affinity Diagram.

Below we present our finalized collective ideas, starting with our definition of HGAI across three levels, followed by detailed discussions of next-steps within each level.